\documentclass[twocolumn]{revtex4}
\usepackage{color}
\usepackage{colordvi}
\usepackage{amsmath}
\usepackage{amssymb}
\usepackage{amsfonts}
\usepackage{psfrag}
\usepackage{epsfig}
\usepackage{bm}
\newcommand{\be}{\begin{equation}}
\newcommand{\ee}{\end{equation}}
\newcommand{\rv}{\mathbf{r}}

\newcommand{\Lv}{\mathbf{L}}
\newcommand{\kv}{\mathbf{k}}

\newcommand{\Gv}{\mathbf{G}}

\begin{document}

\title{The Finite Size Error in Many-body Simulations with Long-Ranged Interactions}

\author{Simone Chiesa}\email{chiesa@uiuc.edu}
\affiliation{ Dept. of Physics, University of Illinois
Urbana-Champaign, Urbana, IL 61801}
\author{D. M. Ceperley}\email{ceperley@uiuc.edu}
\affiliation{ Dept. of Physics, University of Illinois
Urbana-Champaign, Urbana, IL 61801}
\author{R. M. Martin}\email{rmartin@uiuc.edu}
\affiliation{ Dept. of Physics, University of Illinois
Urbana-Champaign, Urbana, IL 61801}
\author{Markus Holzmann}\email{markus@lptl.jussieu.fr}
\affiliation{LPTMC, UMR 7600 of CNRS, Universit{\'e} P. et M. Curie, Paris, France}

\begin{abstract}
We discuss the origin of the finite size error of the energy in many-body
simulation of systems of charged particles and we propose a correction based on
the random phase approximation at long wave
lengths. 
The correction comes from contributions
mainly determined by the organized collective oscillations of the
interacting system. 
Finite size corrections, both on kinetic and potential energy, can be calculated
within a single simulation. 
Results are presented for the electron gas and
silicon. 
\end{abstract}

 \maketitle
The accurate calculation of properties of systems containing
electrons is a very active field of research. Among the possible
numerical approaches, quantum Monte Carlo methods are unique in
their ability to produce reliable ground state properties at a
reasonable computational cost\cite{QMC-RMP}. However, in the simulation of bulk
systems, calculations are necessarily performed using a finite number 
of electrons with a consequent loss in accuracy. In order to reduce 
this bias, 
the system and, hence, the pair interaction are made periodic in a supercell
with basis vectors $\{\Lv_\alpha\}_{\alpha=1,2,3}$. (In the case of a
crystal these vectors define a supercell of the unit cell.) This is
achieved by using the Fourier components of the interaction
at the reciprocal wave vectors of the supercell, i.e. $\Gv$ such
that $\exp(i\Gv\Lv_\alpha)=1$. Singular long-ranged potentials, 
such as the Coulomb interaction, are computed by splitting the sum into a
portion in real and reciprocal space\cite{Natoli}.
Although using the periodized potential reduces the finite-size
effects, some error still remains; the one on the energy, for example, often
exceeds the statistical noise and other errors characteristic of
quantum simulations\cite{footnote1}. Finite size scaling is possible,
but difficult, because the cost of a simulation increases rapidly
with the number of particles in the supercell. Here we present an
approach that reduces the finite size errors.

In Fourier space and atomic units, the electron-electron potential is:
\begin{equation}
V_{e-e}=\frac{2\pi e^2}{\Omega}\sum_{\Gv\ne 0} \frac{1}{G^2}(\rho_\Gv \rho_{-\Gv}-N)
\label{potop}
\end{equation}
where $\rho_\Gv\equiv\sum_i \exp(i\Gv\rv_i)$ is the
Fourier transform of the charge density, $\Omega$ the volume
of the supercell and $N$ the total number of particles. 
The boundary conditions on the wave function can
be chosen as $\Psi(..\rv_i+\Lv_\alpha ..)=
\exp(i\theta_\alpha)\Psi(..\rv_i..)$ where $\theta_\alpha$ is the
``twist'' of the phase in the $\alpha^\text{th}$ direction. Periodic
boundary conditions have $\theta_\alpha=0$. When there is no long
range order, finite size errors are reduced by averaging over
twists (i.e. k-point sampling or Brillioun zone integration)\cite{Lin}. This
comes at little cost in simulations since the average is also
effective in reducing the statistical noise. Even when this is
done, the expectation value of the potential energy remains
expressed as a series over $\Gv$ vectors and is determined by the
static structure factor $S_N(\Gv)=\langle
\rho_\mathbf{G}\rho_\mathbf{-G}\rangle/N$. As the system size increases, the mesh
of $\Gv$ vectors gets finer and the series eventually converges to
an integral corresponding to the exact thermodynamic limit. 

The error using a simulation box with $N$ particles is therefore given by
\begin{equation}
\Delta V\equiv\frac{e^2}{4\pi^2}\int \frac{S_\infty(\kv)-1}{k^2} d\kv -
\frac{2\pi e^2}{\Omega}\sum_{\Gv\ne 0} \frac{S_N(\Gv)-1}{G^2}.
\label{mainerror}
\end{equation}
Its leading order contribution is given by the
Madelung constant, $v_M$, and corresponds to the difference 
$-e^2\int
(2\pi k)^{-2} d\kv + 2\pi e^2\Omega^{-1}\sum_{\Gv\ne 0}G^{-2}$. 
It scales as $1/L$
because of the omission of the
$G=0$ contribution from the sum and its value is proportional to $e^2\int_\mathcal{D}
(2\pi k)^{-2}d\kv$ where $\mathcal{D}$ is a domain of volume
$(2\pi)^3/\Omega$.

Although $v_M$ is generally introduced using a
real space picture, as the interaction between images, the above
perspective can be easily generalized to the next order
correction. The remaining part of the error is determined by i)
the substitution of $S_\infty(\kv)$ by the computed $S_N(\Gv)$ and
ii) the discretization of the integral of $e^2S(\kv)(4\pi^2k^2)^{-1}$. 
The behavior of
$S(\kv)$ at large $k$ is determined by the short range correlation
and can be neglected.
This is apparent if the potential
is decomposed in a short and long range part. The long range part,
whose expectation value is affected by the finite size, decays quickly to $0$
in reciprocal space so that the behavior of $S(\kv)$ at large
$k$ is irrelevant. Moreover, in the limit $k\rightarrow 0$, one
knows that the random-phase approximation becomes exact and
describes independent density-fluctuation modes\cite{Pines_book}.
In the small $k$ region the random-phase approximation 
suggests
\begin{equation}
S_\infty(\Gv)\simeq S_N(\Gv) 
\label{main_hypo}
\end{equation}
and implies that the leading order contribution to the error comes
from point ii) above. It is an integration error that, analogously
to the Madelung constant, comes from the omission of the $\Gv=0$
volume element from the energy sum. Scaling of the finite size
errors is then determined to leading order by this missing
contribution {\em i.e.} $e^2\int_\mathcal{D} S(\kv)(4\pi k^2)^{-1}d\kv$
where $\mathcal{D}$ is a domain centered on $\kv=0$ whose volume
equals $(2\pi)^3/\Omega$. This, together with the
characteristic quadratic behavior of $S(\kv)$ for correlated
charged systems, leads straightforwardly to the well known
$1/\Omega$ scaling of the error. \cite{footnote2} Thanks to the validity of the
random-phase approximation, $S(\kv)$ can be determined in the
small-$k$ region either analytically or from a knowledge of the
$S_N(\kv)$ computed in the simulation. Once $S(\kv)$ is known, one
can accurately compute the correction.

\begin{figure}
\begin{center}
\epsfig{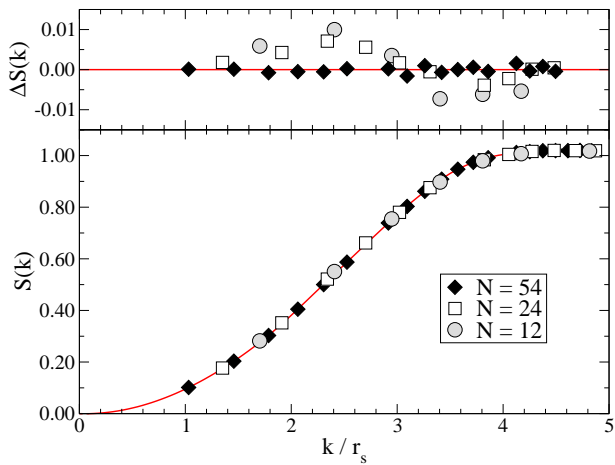} \caption{
Lower panel: Static structure factor for the electron gas at
$r_s=10$. Upper panel: $\Delta S=S_N(k)-S_{66}(k)$. The difference
is computed using a spline function interpolation of $S_{66}$. }
\label{jelliumSk}
\end{center}
\end{figure}

We looked at jellium as a test case to judge to what extent Eq.\ref{main_hypo} 
is verified. Results for $S_N(k)$ computed in variational Monte
Carlo simulations at $r_s=10$ for 12, 24 and 54 particles
are shown in Fig.\ref{jelliumSk}.
As we increase the number of particles, the grid of $k$ points for
which $S_N$ is defined shifts, but the values of $S_N$ fall on a
smooth curve, independent of $N$.  

Let us now consider the kinetic energy. It is important to
distinguish between the effects due to momentum quantization and
long range correlation.  When using a twisted boundary
condition $\bm{\theta}$ in a cubic cell, the kinetic 
energy is given in terms of the momentum distribution by
\begin{equation}
T=\frac{\hbar^2}{2m}\sum_\Gv n_N(\Gv+\bm{\theta}/L) (\Gv+\bm{\theta}/L)^2 \label{nk_finite}
\end{equation}
When using a single twist, for example periodic boundary conditions, 
the finite size error
is, once again, composed of two contributions: the integration
error and the error in approximating the exact momentum
distribution, $n_\infty$, with $n_N$. To better understand the
latter point, consider the fourier transform of the momentum
distribution: the one-body density matrix. This is equal 
to the integral over particle coordinates of
$\Psi^\dagger(\rv_1 + \rv,\rv_2 ...)\Psi(\rv_1,\rv_2...)$ and converges to the exact one as soon as
the correlation length is less than the size of the simulation
box. Under the assumptions of no long range correlation, this
criterion is eventually met so one has $n_N(\kv)=n_\infty(\kv)$
and the error comes again from approximating the thermodynamic
integral with a sum. At variance with the potential energy case, a
change in the twist modifies the grid over which the kinetic
energy is computed (see Eq.\ref{nk_finite}) so that the error can be made
arbitrarily small by increasing the density of twist angles. One
can get away with a small number of special $k$-point in the case
of semiconductors\cite{Rajag1} but a finer grid is needed for a Fermi
liquid due to the discontinuity at the Fermi surface. 
In the latter case the occupation of the single-particle states
varies with the twist and one can use the grand-canonical ensemble
to eliminate this source of error\cite{Lin}.

Consider now the effects due to long range correlation. In Coulomb
systems the interaction causes the wave function to have a
charge-charge correlation factor: the Jastrow potential. Within the
random phase approximation the ground state of the system is described by a collection of
dressed particles interacting via short range forces and quantized
coherent modes, the plasmons. Accordingly, the many-body wave function
factorizes as\cite{Bohm}
\begin{equation}
\Psi=\Psi_\text{s.r.} \exp\left[-\frac{1}{2\Omega} \sum_{\Gv\ne 0} u_\Gv\rho_\Gv\rho_\Gv^\dagger\right]
\end{equation}
where $\Psi_\text{s.r.}$ only contains short range correlations
and $u_\Gv$ decays quickly to $0$ as $G$ increases and diverges as
$G^{-2}$ at small $G$. Because of this divergence, $n_N$
converges very slowly to its thermodynamic value and the average
over twists provides only a partial correction. Although one can
address the bias on the momentum distribution\cite{Magro} directly,
we here employ a different route.
Thanks to Green's identity the
kinetic energy is written as
$T=-\left\langle\hbar^2\nabla^2\ln\psi\right/4m\rangle$\cite{McMillan} 
with a contribution coming from the Jastrow potential given by
\begin{equation}
T_N= -\frac{\hbar^2}{4m \Omega} \sum_{\Gv\ne 0} G^2 u_\Gv \left[ S(\Gv) -1 \right].
\label{TNB}
\end{equation}
Hence the error of the kinetic energy also has the form
of Eq.\ref{mainerror}:
a $1/\Omega$ finite size error in the kinetic energy
corresponding to the omission of the $\Gv=0$ term in Eq.\ref{TNB}\cite{footnote3}. This is an
integration error provided $u_k$ does not depend on the system
size. This must be the case whenever Eq.\ref{main_hypo} is satisfied since a
difference in $u_k$ would necessarily imply a difference in
$S(\kv)$.

Errors in the potential and kinetic energy have therefore a very similar mathematical structure.
To compute the two corrections we use the Poisson sum formula
$\sum_{\Lv} \widetilde{\zeta}(\Lv)=\Omega^{-1}\sum_{\Gv} \zeta(\Gv)$ where
$\widetilde{\zeta}$ and $\zeta$ are a Fourier transform pair. By
separating the $\Lv=0$ and $\Gv=0$ contributions from the two sums
we get the expression for the error
\begin{equation}
\Delta_N\equiv\frac{1}{(2\pi)^3}\int \zeta(\kv)d\kv-\sum_{\Gv\ne 0} 
\frac{\zeta(\Gv)}{\Omega}=
\frac{\zeta(0)}{\Omega}-\sum_{\Lv\ne0}\widetilde{\zeta}(\Lv).
\label{PSF}
\end{equation}
One sets
$\zeta(0)$ equal to the $k=0$ limit of $2\pi e^2S(k)k^{-2}$ or $\hbar^2k^2u(k)/4m$ 
for the correction to the potential and kinetic energy respectively.

\begin{figure}
\begin{center}
\epsfig{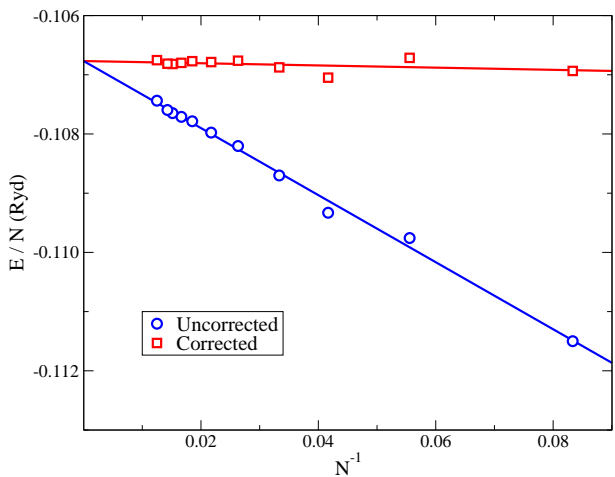}
\caption{Energies per particle of the electron gas at $r_s=10$
in Rydberg as a function of the inverse particle number. Circles are
the Monte Carlo energies averaged over twist angles. Squares are the
energies after the additional $\hbar\omega_p/2 N$ correction (see text).
}
\label{jellium_summary}
\end{center}
\end{figure}

We first apply these corrections to the electron gas for which the
small $k$ limits of $S(k)$ and $u(k)$ are known from
the random phase approximation as, respectively, $\hbar
k^2/2m\omega_p$ and $4\pi e^2/\hbar\omega_pk^2$ where $\omega_p$
is the plasma frequency.
In our tests, the wave function had a
backflow-Jastrow form\cite{Holzmann} and simulations were performed
in the grand-canonical ensemble. Thanks to 
the translational invariance of the Hamiltonian, the wave function factorizes as
$\exp(i\bm{\theta}\sum_i \rv_i/L)\Phi$
where $\Phi$, the periodic part, is invariant in a finite pocket of $k$-space around each twist
angle. In each pocket the energy dependence on $\bm{\theta}$ is trivial
and one can exploit this fact to reduce the number of twist angles to be the number of
inequivalent pockets. This, together with cubic symmetry,
drastically reduces the
number of needed twist angles to between $20-200$
for an unpolarized system with $N\sim 10 - 100$.
The leading order correction due to long range correlations
to kinetic and
potential energy are equal and sum up a total error
$\Delta_N=\hbar\omega_p(2N)^{-1}$. Corrected and uncorrected
variational energies are shown in Fig.\ref{jellium_summary} for
$r_s=10$. Diffusion Monte Carlo values are uniformly shifted to lower energy by 0.6
mRyd/electron and show similar behavior. One can see that the
bias due to the small size of the simulation cell is tremendously
reduced, so that the $N=12$ case is already satisfactory.

\begin{figure}
\begin{center}
\epsfig{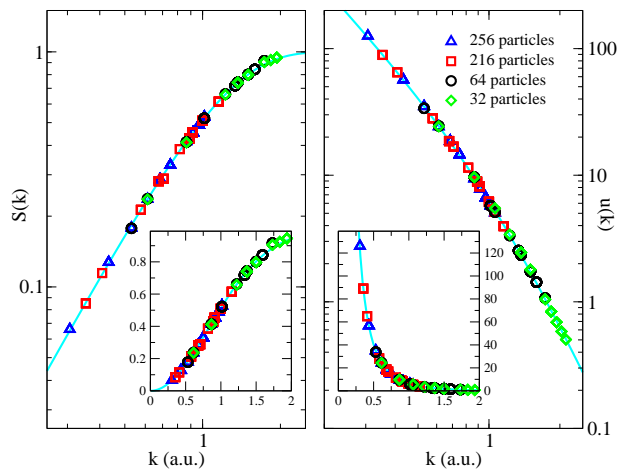}
\caption{Structure factor (left) and Jastrow potential (right)
for diamond silicon at ambient pressure. The continuous lines are fit
to the data (see text). The Jastrow potential shows a $k^{-2}$ divergence 
at small $k$ that was not explicitely imposed but obtained through
energy variance minimization using the \textsc{casino} code.}
\label{SiliconSK}
\end{center}
\end{figure}
As a second example we considered the diamond structure of silicon
at ambient pressure ($r_s=2.0$). Calculations were
performed using the \textsc{casino}\cite{CASINO} code, a Slater-Jastrow
wave function, a Hartree-Fock pseudopotential\cite{Trail1,Trail2} 
and periodic boundary conditions. The orbitals
used for the trial function (Hartree-Fock) were from the \textsc{crystal}{\footnotesize 98}
code\cite{Crys98}. To eliminate the effects of momentum
quantization we used a correction based
on the density functional eigenvalues of those single-particle states 
periodic in the simulation cell. Although this is quite common practice 
it involves another
uncontrolled approximation and results depend weakly on the
functional employed (we used the local density approximation).
The parameters in the Jastrow potential and a one-body term were
optimized. The two-particle Jastrow factor was made up by
a spherical short range part and a plane wave expansion
including 3 shells of k-points\cite{DrummondJas}. One needs the plane wave expansion
to accurately reproduce the behavior of the exact Jastrow factor at 
small $k$, especially in the case of small simulation cells. To further
eliminate errors in the wave function we correct the diffusion Monte Carlo value of
$S(k)$ by $S_\Gv^{\text{DMC}}-S_\Gv^{\text{VMC}}$ which leads to an
estimate correct to second-order in the wave function.

\begin{figure}
\epsfig{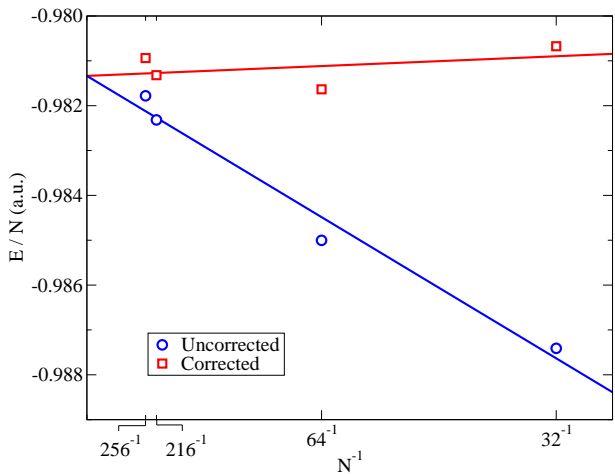}
\caption{Diffusion Monte Carlo energies per electron in diamond Silicon at 
$r_s=2.0$. Energies and the $S(k)$ and $u(k)$ used to compute the
correction are all obtained in simulations with the same number of particles.
The smallest cell is the conventional fcc cubic cell of diamond.
The two intermidiate ones are, respectively, $2\times2\times2$ and $3\times3\times3$ supercells of the 
primitive cell. The largest one is a $2\times2\times2$ supercell of the conventional cubic cell.}
\label{Si_summary}
\end{figure}

For Eq.\ref{PSF} we assumed $S(k)=1-\exp(-\alpha k^2)$ and
$u(k)=4\pi a[k^{-2}-(k^2+a^{-1})^{-1}]$\cite{Becker}.
When $k$ is expressed
in atomic units, the optimal value of 
$\alpha$ and $a$ were found to be $0.72$ and $1.0$ respectively, leading to
corrections of $0.13/N$ and $0.092/N$ hartree per electron for
potential and kinetic energy.
Results after the two corrections were applied are shown in 
Fig.\ref{Si_summary}. Even for the smallest cell (cubic, 
with 8 Si atoms), the error in the energy is of the order of 
1 mHartree/electron (0.1 eV/atom) 
when compared to the value extrapolated for the infinite size.

To conclude, we propose a way to estimate the errors in the
potential and kinetic energy under the assumption that the low $k$
behavior of the correlation factor is unchanged upon variation of
the simulation cell size. This scheme is suggested by the
random-phase approximation that describes independent collective
mode in the limit $k\rightarrow 0$. The dominant finite size errors 
on potential and
kinetic energy are integration errors that can be estimated by
using the properties of the charge structure factor and the
Jastrow potential at long wavelength. 
The behavior of these
quantities in the small $k$ limit can either be obtained analytically (e.g. for the
electron gas) or from results with accurate optimized trial wave
functions.  This approach can be used to obtain energies close to
the thermodynamic limit without performing a scaling analysis
using different sized systems or assuming the finite-size behavior is given
by Fermi liquid theory or approximated by density functional theory.

This material is based upon work supported in part by
the U.S Army Research Office under DAAD19-02-1-0176.
Computational support was provided by the Materials
Computational Center (NSF DMR-03 25939 ITR), the National Center for
Supercomputing Applications at the University of
Illinois at Urbana-Champaign and by CNRS-IDRIS.


\begin{thebibliography}{18}
\expandafter\ifx\csname natexlab\endcsname\relax\def\natexlab#1{#1}\fi
\expandafter\ifx\csname bibnamefont\endcsname\relax
  \def\bibnamefont#1{#1}\fi
\expandafter\ifx\csname bibfnamefont\endcsname\relax
  \def\bibfnamefont#1{#1}\fi
\expandafter\ifx\csname citenamefont\endcsname\relax
  \def\citenamefont#1{#1}\fi
\expandafter\ifx\csname url\endcsname\relax
  \def\url#1{\texttt{#1}}\fi
\expandafter\ifx\csname urlprefix\endcsname\relax\def\urlprefix{URL }\fi
\providecommand{\bibinfo}[2]{#2}
\providecommand{\eprint}[2][]{\url{#2}}

\bibitem[{\citenamefont{Foulkes et~al.}(2001)\citenamefont{Foulkes, Mitas,
  Needs, and Rajagopal}}]{QMC-RMP}
\bibinfo{author}{\bibfnamefont{W.~M.~C.} \bibnamefont{Foulkes}},
  \bibinfo{author}{\bibfnamefont{L.}~\bibnamefont{Mitas}},
  \bibinfo{author}{\bibfnamefont{R.~J.} \bibnamefont{Needs}}, \bibnamefont{and}
  \bibinfo{author}{\bibfnamefont{G.}~\bibnamefont{Rajagopal}},
  \bibinfo{journal}{Rev. Mod. Phys.} \textbf{\bibinfo{volume}{73}},
  \bibinfo{pages}{33} (\bibinfo{year}{2001}).

\bibitem[{\citenamefont{Natoli and Ceperley}(1995)}]{Natoli}
\bibinfo{author}{\bibfnamefont{V.}~\bibnamefont{Natoli}} \bibnamefont{and}
  \bibinfo{author}{\bibfnamefont{D.~M.} \bibnamefont{Ceperley}},
  \bibinfo{journal}{J. Comp. Phys.} \textbf{\bibinfo{volume}{117}},
  \bibinfo{pages}{171} (\bibinfo{year}{1995}).

\bibitem[{foo({\natexlab{a}})}]{footnote1}
\bibinfo{note}{For example, the time-step bias and the fixed node error.}

\bibitem[{\citenamefont{Lin et~al.}(2001)\citenamefont{Lin, Zong, and
  Ceperley}}]{Lin}
\bibinfo{author}{\bibfnamefont{C.}~\bibnamefont{Lin}},
  \bibinfo{author}{\bibfnamefont{F.~H.} \bibnamefont{Zong}}, \bibnamefont{and}
  \bibinfo{author}{\bibfnamefont{D.~M.} \bibnamefont{Ceperley}},
  \bibinfo{journal}{Phys. Rev. E} \textbf{\bibinfo{volume}{64}},
  \bibinfo{pages}{16702} (\bibinfo{year}{2001}).

\bibitem[{\citenamefont{Nozieres and Pines}(1999)}]{Pines_book}
\bibinfo{author}{\bibfnamefont{P.}~\bibnamefont{Nozieres}} \bibnamefont{and}
  \bibinfo{author}{\bibfnamefont{D.}~\bibnamefont{Pines}},
  \emph{\bibinfo{title}{The theory of quantum Liquids}}
  (\bibinfo{publisher}{Perseus Books}, \bibinfo{year}{1999}).

\bibitem[{foo({\natexlab{b}})}]{footnote2}
\bibinfo{note}{Note, however, that in the Hartree-Fock approximation, metallic
  systems are characterized by $S(\kv)\propto k$ at small $k$ so that the error
  is expected to scale as $\Omega^{-2/3}$.}

\bibitem[{\citenamefont{Rajagopal et~al.}(1994)\citenamefont{Rajagopal, Needs,
  Kenny, Foulkes, and James}}]{Rajag1}
\bibinfo{author}{\bibfnamefont{G.}~\bibnamefont{Rajagopal}},
  \bibinfo{author}{\bibfnamefont{R.~J.} \bibnamefont{Needs}},
  \bibinfo{author}{\bibfnamefont{S.}~\bibnamefont{Kenny}},
  \bibinfo{author}{\bibfnamefont{W.~M.~C.} \bibnamefont{Foulkes}},
  \bibnamefont{and} \bibinfo{author}{\bibfnamefont{A.}~\bibnamefont{James}},
  \bibinfo{journal}{Phys. Rev. Lett.} \textbf{\bibinfo{volume}{73}},
  \bibinfo{pages}{1959} (\bibinfo{year}{1994}).

\bibitem[{\citenamefont{Bohm and Pines}(1953)}]{Bohm}
\bibinfo{author}{\bibfnamefont{D.}~\bibnamefont{Bohm}} \bibnamefont{and}
  \bibinfo{author}{\bibfnamefont{D.}~\bibnamefont{Pines}},
  \bibinfo{journal}{Phys. Rev.} \textbf{\bibinfo{volume}{92}},
  \bibinfo{pages}{609} (\bibinfo{year}{1953}).

\bibitem[{\citenamefont{Magro and Ceperley}(1994)}]{Magro}
\bibinfo{author}{\bibfnamefont{W.~R.} \bibnamefont{Magro}} \bibnamefont{and}
  \bibinfo{author}{\bibfnamefont{D.~M.} \bibnamefont{Ceperley}},
  \bibinfo{journal}{Phys. Rev. Lett.} \textbf{\bibinfo{volume}{73}},
  \bibinfo{pages}{826} (\bibinfo{year}{1994}).

\bibitem[{\citenamefont{McMillan}(1965)}]{McMillan}
\bibinfo{author}{\bibfnamefont{W.~L.} \bibnamefont{McMillan}},
  \bibinfo{journal}{Phys. Rev.} \textbf{\bibinfo{volume}{138}},
  \bibinfo{pages}{A442} (\bibinfo{year}{1965}).

\bibitem[{foo({\natexlab{c}})}]{footnote3}
\bibinfo{note}{The dominant contribution is given by the ``$-1$''. The form of
  the leading order error is $\hbar^2[4m (2\pi)^3]^{-1}\int d\kv k^{2}u(\kv) -
  \hbar^2(4m \Omega)^{-1} \sum_{\Gv\ne 0} G^2 u_\Gv $.}

\bibitem[{\citenamefont{Holzmann et~al.}(2003)\citenamefont{Holzmann, Ceperley,
  Pierleoni, and Esler}}]{Holzmann}
\bibinfo{author}{\bibfnamefont{M.}~\bibnamefont{Holzmann}},
  \bibinfo{author}{\bibfnamefont{D.~M.} \bibnamefont{Ceperley}},
  \bibinfo{author}{\bibfnamefont{C.}~\bibnamefont{Pierleoni}},
  \bibnamefont{and} \bibinfo{author}{\bibfnamefont{K.}~\bibnamefont{Esler}},
  \bibinfo{journal}{Phys. Rev. E} \textbf{\bibinfo{volume}{68}},
  \bibinfo{pages}{46707} (\bibinfo{year}{2003}).

\bibitem[{\citenamefont{Needs et~al.}(2004)\citenamefont{Needs, Towler,
  Drummond, and Kent}}]{CASINO}
\bibinfo{author}{\bibfnamefont{R.}~\bibnamefont{Needs}},
  \bibinfo{author}{\bibfnamefont{M.}~\bibnamefont{Towler}},
  \bibinfo{author}{\bibfnamefont{N.}~\bibnamefont{Drummond}}, \bibnamefont{and}
  \bibinfo{author}{\bibfnamefont{P.}~\bibnamefont{Kent}},
  \emph{\bibinfo{title}{CASINO version 1.7 User Manual}}
  (\bibinfo{publisher}{University of Cambridge, Cambridge},
  \bibinfo{year}{2004}).

\bibitem[{\citenamefont{Trail and Needs}(2005{\natexlab{a}})}]{Trail1}
\bibinfo{author}{\bibfnamefont{J.~R.} \bibnamefont{Trail}} \bibnamefont{and}
  \bibinfo{author}{\bibfnamefont{R.~J.} \bibnamefont{Needs}},
  \bibinfo{journal}{J. Chem. Phys.} \textbf{\bibinfo{volume}{122}},
  \bibinfo{pages}{174109} (\bibinfo{year}{2005}{\natexlab{a}}).

\bibitem[{\citenamefont{Trail and Needs}(2005{\natexlab{b}})}]{Trail2}
\bibinfo{author}{\bibfnamefont{J.~R.} \bibnamefont{Trail}} \bibnamefont{and}
  \bibinfo{author}{\bibfnamefont{R.~J.} \bibnamefont{Needs}},
  \bibinfo{journal}{J. Chem. Phys.} \textbf{\bibinfo{volume}{122}},
  \bibinfo{pages}{014112} (\bibinfo{year}{2005}{\natexlab{b}}).

\bibitem[{\citenamefont{Saunders et~al.}(1998)\citenamefont{Saunders, Dovesi,
  Roetti, Causa, Harrison, Orlando, and Zicovich}}]{Crys98}
\bibinfo{author}{\bibfnamefont{V.~R.} \bibnamefont{Saunders}},
  \bibinfo{author}{\bibfnamefont{R.}~\bibnamefont{Dovesi}},
  \bibinfo{author}{\bibfnamefont{C.}~\bibnamefont{Roetti}},
  \bibinfo{author}{\bibfnamefont{M.}~\bibnamefont{Causa}},
  \bibinfo{author}{\bibfnamefont{N.~M.} \bibnamefont{Harrison}},
  \bibinfo{author}{\bibfnamefont{R.}~\bibnamefont{Orlando}}, \bibnamefont{and}
  \bibinfo{author}{\bibfnamefont{C.~M.} \bibnamefont{Zicovich}},
  \emph{\bibinfo{title}{CRYSTAL98 User's Manual}}
  (\bibinfo{publisher}{University if Torino}, \bibinfo{year}{1998}).

\bibitem[{\citenamefont{Drummond et~al.}(2004)\citenamefont{Drummond, Towler,
  and Needs}}]{DrummondJas}
\bibinfo{author}{\bibfnamefont{N.~D.} \bibnamefont{Drummond}},
  \bibinfo{author}{\bibfnamefont{M.~D.} \bibnamefont{Towler}},
  \bibnamefont{and} \bibinfo{author}{\bibfnamefont{R.~J.} \bibnamefont{Needs}},
  \bibinfo{journal}{Phys. Rev. B} \textbf{\bibinfo{volume}{70}},
  \bibinfo{pages}{235119} (\bibinfo{year}{2004}).

\bibitem[{\citenamefont{Becker et~al.}(1968)\citenamefont{Becker, Broyles, and
  Dunn}}]{Becker}
\bibinfo{author}{\bibfnamefont{M.~S.} \bibnamefont{Becker}},
  \bibinfo{author}{\bibfnamefont{A.~A.} \bibnamefont{Broyles}},
  \bibnamefont{and} \bibinfo{author}{\bibfnamefont{T.}~\bibnamefont{Dunn}},
  \bibinfo{journal}{Phys. Rev.} \textbf{\bibinfo{volume}{175}},
  \bibinfo{pages}{224} (\bibinfo{year}{1968}).

\end{thebibliography}
\end{document}